\documentstyle[12pt]{article}

\newcommand{\bc}{{\bf C}}

\newcommand{\la}{\mbox{$\lambda$}}

\newcommand{\ra}{\mbox{$\rightarrow $}}

\newcommand{\pa }{\partial }

\newcommand{\Pf}{\noindent {\it Proof:}\ }
\newcommand{\ep}{\epsilon}

\newcommand{\al}{\alpha }

\newcommand{\bR}{{\bf R}}
\newcommand{\bC}{{\bf C}}
\newcommand{\bo}{{\cal L}(L^2(\Omega )) }
\newcommand{\tR}{\tilde{S}}
\newtheorem{lemma}{Lemma}
\newtheorem{prop}{Proposition}
\newtheorem{thm}{Theorem}
\newtheorem{cor}{Corollary}
\title{On the resonances of the Laplacian on waveguides}

\author{Julian Edward\footnotemark[2]}

\begin{document}
\maketitle

\renewcommand{\thefootnote}{\fnsymbol{footnote}}

\footnotetext[2]{
Department of Mathematics,
Florida International Univ.,
Miami, FL 33199,
edwardj@fiu.edu}

\renewcommand{\thefootnote}{\arabic{footnote}}

{\bf AMS Classification numbers:} 35P25, 81U99, 76Q05, 58J50.

{\bf Keywords:} Resonances, Laplacian, Waveguide.

%Send proofs to: Julian Edward at address on first page.

\begin{abstract}
The resonances for the Dirichlet and Neumann Laplacian
are studied on compactly perturbed waveguides.
In the absence of resonances, an
upper bound is proven for the localised resolvent.
This is then used to prove that the existence of a
quasimode whose asymptotics is bounded away from
the thresholds implies the existence of resonances
converging to the real axis. The following upper bound to the number of resonances
is also proven:
$$\# \{ k_j\in Res (\Delta ),\ dist(k_j,\mbox{physical plane})<1+\sqrt{|k_j|}/2
, |k_j|<r \} <Cr^{3+\ep}.$$
\end{abstract}

\markboth{J.K. EDWARD }{RESONANCES OF THE LAPLACIAN}

\begin{section}{Introduction}

Resonances of the Laplacian have been the object of study in
a wide variety of geometric settings (see surveys in \cite{Z,Z4,V3}).
Resonances, which are essentially equivalent to poles of the
scattering matrix, have been related to long-lived waves
(``metastable states'' in the quantum mechanics literature) and
also arise naturally in studying the long time
behaviour of evolution equations, particularly the wave
equation (see \cite{SW}, where this connection is pointed
out in the context of Schro$\ddot{d}$inger operators).

Despite the physical significance of resonances, very little is
understood about the resonances associated to perturbations
of waveguides. The only work 
known to this author that explicitly studies resonances is by Aslanyan-Parnovski-Vassiliev \cite{apv}, where the authors estimate the complex
part of resonances arising as perturbations of trapped modes. The authors
also use numerical methods to count the resonances that appear at low
frequencies. 

Several other works on the scattering theory of 
waveguides indirectly apply to resonances.
Christiansen-Zworski \cite{CZ}, and independantly Parnovksi \cite{P}, have
computed the asymptotics for the embedded eigenvalue counting function added
to the scattering phase for boundariless manifolds  which are asymptotic
to a cylinder. They also proved a sharp upper bound on the eigenvalue
counting function, improving on an earlier work by Donnelly  \cite{Do}.
Melrose \cite{M} studied the properties of the resolvent
for asymptotically perturbed cylinders, including the nature of the 
singularity of the resolvent at the thresholds. Weidenmuller \cite{W}
studied various scattering theoretic properties of Laplacian on
the perturbed strip with Dirichlet boundary conditions.

Also related  is the extensive literature on the existence of 
$L^2$ eigenvalues.
With our definition of resonance, any $L^2$-eigenvalue will be
considered part of the resonance set. The existence of
$L^2$-eigenvalues for waveguides has drawn much study,
motivated both by their association with standing waves
or ``trapped modes'' (see \cite{Va,E} and references therein)
and with their role in forcing equations (\cite{W2,W3} and references
therein). 
However, it is generally believed that $L^2$-eigenvalues
do not exist for generic perturbations of the strip.

In this paper, we consider the resonances associated
to the Laplacian on waveguides with either Dirichlet or Neumann
boundary conditions.
We prove a resolvent estimate
from which it follows that the existence of a quasimode with
certain asymptotics
implies the existence of a sequence of resonances approaching the
real axis. We also prove an upper bound on the number of resonances
in a neighbourhood of the physical plane.

We define resonances as
poles of the meromorphic continuation of the resolvent away from the 
thresholds, in union with any $L^2$ eigenvalues occuring at the thresholds.
In particular,
let $\Omega \subset \bR ^2$ be a domain with smooth boundary
which to the exterior of some compact set coincides with
the strip
$(-\infty ,\infty )\times (0,\pi )$.
Let $\Delta$ be either the Dirichlet or Neumann Laplacian,
with signs arranged so that the operator is positive
semi-definite. It is well known that the  Dirichlet Laplacian has
essential spectrum  $[1,\infty )$,
with thresholds at $\{ j^2\}_{j=1}^{\infty}$.
For the Neumann Laplacian, the essential spectrum is  $[0,\infty )$
with thresholds at $\{ j^2\}_{j=0}^{\infty}$.
Let $\chi \in
C_0^{\infty}({\bf R}^2)$. We show in Section 2 that
$\chi (\Delta -k)^{-1}\chi$, an analytic function in $k$ with
values in the bounded operators  on $L^2(\Omega )$,
extends meromorphically from $\bc -[0,\infty )$ to an infinitely
branched Riemann surface $S$, with the branch points
occurring at the thresholds. 

The geometry of $S$ was studied in \cite{W},
where it is proven that $S$ is not simply connected. This, and 
the infinitely many branch points, are probably the
main reasons that the resonances on waveguides are less well understood
than for the corresponding problem for exterior domains
(for exterior domains the corresponding Riemann surface
is the surface associated to $\sqrt{z}$ for odd dimensions, and
$\ln z$ for even dimensions). Also, the tools
of complex scaling as in \cite{Sj} have not
been established in this setting.

Let $\Pi: S\ra \bc$ be the
canonical projection. The Riemannian metric induced by $\Pi$
naturally induces a distance function on $S$, which we denote $dist$.
We prove the following
resolvent estimate:
\begin{thm}
Let $\Omega \subset \bR^2$ be a smooth domain which to the exterior
of a bounded set equals the strip $\{ (x,y): x\in (-\infty,\infty ),
\ y\in (0,\pi)\} $. Fix $\al >0$. Then for integer $p\geq 3$,
$q\in [0,p/2)$, there exists
 $M_{p,q }>0$ such that if
 $m>M_{p,q}$, the following property holds:
 if there are no
thresholds in the interval $(m-\al-2m^{-q} ,m+\al+2m^{-q} )$
and no resonances
in the open set
$$\{ k\in S: \ dist(k,[m-2m^{-q}
,m+2m^{-q}])< 2(m-2)^{-p}\} , $$
then for $k\in [m-m^{-q},m+m^{-q}]$, we have
$$\| \chi (\Delta -k)^{-1}\chi \|_{L^2\ra L^2}\leq C_pk^{2p}.$$
Here $C_p$ is a positive constant depending only on $p$ and $\Omega$.
\end{thm}
We also prove some upper bounds in a neighbourhood of the thresholds; see
Lemma 7.
A consequence of this theorem is that the existence of
 periodic billiard trajectories in $\Omega$ satisfying certain geometric
hypotheses
will imply the existence of a sequence of resonances
converging to the real axis. To be specific we must first
define localised quasimodes.

We define a pair of sequences
$(u_j,\la_j),$ with $u_j$ in the operator domain of $\Delta$ and
$\la_j\in \bR$, to be a quasimode if
the $u_j$ are uniformly
compactly supported  with $\| u_j\|=1$,
$\la_j\ra \infty$, and
$$
\| (\Delta -\la_j)u_j\|_{L^2(\Omega )} = O(\la_j^{-\infty }).$$
Quasimodes that are supported in a neighbourhood of
a stable periodic billiard trajectory 
and for which the asymptotics of $\la_j$ are fully determined by
the local geometry
 have been constructed by a number of authors
(\cite{L,Bab,Bul}). In Section 4 of this paper, we briefly present an
example due to Buldyrev.

\begin{cor}
Suppose there exists a quasimode
such that there exists $\al >0$ such that
$$|\la_j-n^2|>\al ,\forall n,j\in {\bf Z}.$$
Then
 there exists an infinite sequence $\{ k_j\} $ of resonances
of $\Delta$ such that for any $N>0$,
$$|\la_j-k_j|\leq C_N\la_j^{-N}.$$
\end{cor}
Corollary 1 follows from Theorem 1 by direct application of the arguments
appearing in \cite{TZ}.
Such ``quasimode-to-resonance'' results, based on $apriori$ resolvent
estimates, have previously been proven for other geometric settings,
\cite{SV},\cite{TZ},\cite{St},\cite{st2},\cite{st3}.

If the quasimodes also satisfy certain
spacing hypotheses, then
Theorem 1 would also imply that the
 resonance counting function  is bounded below by the
quasimode counting function (see \cite{TZ}).

Using estimates proven in Theorem 1 along with Jensen's formula,
one can also obtain an upper bound on the number of resonances
near the physical plane.
\begin{thm}
Let $\{ k_j\} $ be the resonances of $\Delta$, counted with multiplicity.
Define
$$N(r)=\{ k_j:\ dist(k_j,\mbox{physical plane})<
1+\frac{1}{2}\sqrt{|k|, \ |k|<r}
 \}.$$ 
Then for any $\ep >0$,
there exists a positive constant $C$ such that 
$$N(r)<Cr^{3+\ep}.$$
\end{thm}
Using the methods of this paper, one could also obtain a global upper
bounds on the number of resonances.

Upper bounds for the number of resonances proven in other geometries
suggest that the sharp upper bound for $N(r)$ should of the form
$Cr^2$. It should also be noted that Christiansen-Zworski in \cite{CZ}
proved that the sharp upper bound for the embedded eigenvalue counting 
function is $Cr^2$.

We now give a sketch of the proof, which is based on the Fredholm
determinant method. Let $\chi_1,\chi_2$
be smooth cutoff functions of bounded support. Then
using a well known procedure (see, eg., \cite{SjZ}),
we show that $\chi_1 (\Delta -k)^{-1}\chi_2$ extends
meromorphically to the Riemann surface $S$. It is well known
that the (non-threshold) poles of the resolvent are among the zeros
of a certain Fredholm determinant which is analytic on $S$ away from
the thresholds.
We use estimates for the Green's function for the unperturbed
strip and  adapt  arguments
previously used to study resonances for the exterior problem
(\cite{M,Z2,V}) to obtain an upper bound on the Fredholm determinant.
Using the minimodulus theorem of Cartan together
with an adaptation of a minimodulus theorem for sectors 
found in \cite{cart}, we
obtain a lower bound on the Fredholm determinant. 

Arguing as
in \cite{Z3}, we then obtain an $apriori$
estimate on the extended resolvent in an open set
away from the thresholds and away from the resonances.
Theorem 1 is then proven using an application of the maximum principle
inspired by one used in \cite{TZ}, where the argument is given in
a semi-classical framework.
The use of the Cartan theorem in the study of spectral and scattering theory  
was initiated in \cite{Ma},\cite{Z9}, 
and the use of the minimodulus result for sectors by \cite{PZ}.
For a different proof of the minimodulus theorem for sectors see
\cite{sj2}. 

Theorem 2  follows by applying
Jensen's formula for meromorphic functions \cite{cart}, together with
the upper and lower bounds on the Fredholm determinant,
to obtain upper bounds on the number of resonances on disks, the union of
which covers the positive real axis. Remark: for the exterion problem
in odd dimensions, Jensen's
formula was used to obtain global, sharp  upper bounds on the number of
resonances \cite{Z2}.

We conclude this section by observing that the methods
used in this paper could easily be applied to prove
analagous results for perturbations of more general
cylinders, in particular the standard cylinder in ${\bf R}^3$:
$$\{ (x,y,z): x^2+y^2<1, \ z\in (-\infty ,\infty )\}. $$

\end{section}
\begin{section}{Preliminaries}
We prove our results for the Neumann Laplacian, leaving it
to the reader to make the simple modifications necessary
for Dirichlet boundary conditions.

Let
$$\Omega_0 =\{ (x,y): x\in (-\infty ,\infty ),\ y\in (0,\pi )\} .$$
Let $\Omega $ be a domain with smooth boundary such that there
exists $M>0$ so that
\begin{equation}
\Omega -\{ \sqrt{x^2+y^2}>M\}
=\Omega_0 -\{ \sqrt{x^2+y^2}>M\}.\label{omega}
\end{equation}
On such a domain, we define the  Neumann
Laplacian, $\Delta$, as
the operator living on $L^2(\Omega )$ with
$$\Delta u\equiv -\frac{\pa ^2u}{\pa x^2}
-\frac{\pa ^2u}{\pa y^2},$$
and with operator core
$$\{ u\in  L^2(\Omega ),\ \Delta u\in L^2(\Omega ),\ \frac{\pa u}{\pa \eta}=0\}.$$
Here $\pa/ \pa \eta$ is the normal derivative at the boundary of
$\Omega$.
(For the Dirichlet Laplacian, the operator core is $C_0^{\infty}(\Omega )$ .)

Denote by $L^2(\Omega )$ the set of square integrable
functions on $\Omega $, and  the set of bounded
operators on $L^2(\Omega )$ by $\bo$.
Denote by $B(a,r)$ the ball centered at $a$ of radius $r$,
intersected with $\Omega$ when appropriate.
Denote the Neumann Laplacian on $\Omega$ (resp. $\Omega_0$) by
$\Delta$ (resp. $\Delta_0$).
Define the Sobolev spaces $H^i(\Omega )$ as the
operator domains of $(\Delta +1)^{i/2}$.
We
define a smooth partition of unity $\chi_1+\chi_2=1$
such that $\chi_i \geq 0$,
 $supp(\chi_1 )\subset B(0,M+2)$,
and $\chi_1 =1$ on $B(0,M+1)$. We also define smooth cutoff functions
$\tau_i \geq 0$ such that $\tau_1 =1$ on $supp (\chi_1)$ and
$supp(\tau_1 )\subset B(0,M+3)$, and
$\tau_2 =1$ on $supp (\chi_2)$ and
$\tau_2 =0$ on $ B(0,M)$.
Finally, we define a smooth cutoff function
$\rho$
 such that the $supp(\rho)\subset B(0,M+4)$ and
\begin{equation}
\rho|_{B(0,M+3)}=1.
\label{rho}
\end{equation}

Denote
the associated resolvent $(\Delta_0-k)^{-1}$ by $R_0(k)$.
Denote $(\Delta-k)^{-1}$ by $R(k)$.
Then it is well known that
the Green function for  the operator $\Delta_0 -k$, ie. the
Schwartz kernel for $R_0(k)$, is given by
\begin{equation}
G_k (x,y,x',y')=\frac{1}{\pi \sqrt{-k}}e^{-\sqrt{-k}|x-x'|}
+\sum _{n=1}^{\infty}
\frac{1}{\sqrt{n^2-k}}
e^{-\sqrt{n^2-k}|x-x'|}\cos (ny) \cos (ny').\label{green}
\end{equation}
In the formula above, the branch lines for the
functions $k\ra \sqrt{n^2-k}$
are assumed
to be $[n^2 ,\infty )$.
In what follows,
let $\arg_n$ be the argument associated to the branch
point $n^2$. For each square-root function,
the set $\{ k: \arg_n (k)\in (0,2\pi)\} $ will be
referred to as the ``physical branch'', and the set
$\{ k: \arg_n (k)\in [2\pi ,0]\}$ the non-physical
branch.

Let $S$ be
the infinitely branched Riemann surface
 associated to $G_k$. Thus $G_k$ extends pointwise to $S$.
The geometry of $S$ was studied in \cite{W}.
A point $k\in S$ will be
on the ``physical branch of $S$'' if
$$\arg_n (k)\in (0,2\pi ), \ \forall n;$$
thus the physical branch of $S$ can be identified with
the subset $\bC -[0,\infty )$ of the complex plane.
For $k\in S$, we denote by $\Lambda (k)$
the finite multi-index $(n_1,\ldots ,n_P)$ such that
$k$ is on the ``non-physical branch'' precisely for the
functions $\{ \sqrt{n_i^2-k} \}_{i=1}^P$.

Let $\Pi$ be the
canonical projection of $S$ onto ${\bf C}$.
The induced Riemannian metric on $S$ induces a distance function
that we will denote $dist$.
It will also be convenient to define the
following functions on $S$: $|k|\equiv |\Pi k|$,
$\Re k\equiv \Re \Pi k$, and $\Im k\equiv \Im \Pi k$.
Note that since $\Pi$ is not a global isometry,  $dist(z_1,z_2)$ is not
always equal to $|z_1-z_2|$.

Since we will be concerned with the behaviour of the resolvent only in
a  neighbourhood of the physical plane, we shall define the following
subsets of $S$.
$$
\tilde{S}=
\{ k\in S: dist(k,\mbox{physical plane})<1+\sqrt{|k|} \} .
$$
It will also be convenient to study $\tilde{S}$ away from the thresholds, 
hence for $\al >0$,
$$\tilde{S}_{\alpha}=\{ k\in \tilde{S},
 dist(k,n^2)>\al , \ \forall n\in {\bf Z} \} .$$

For $k\in \tilde{S}$, define $n_k$ to the the greatest integer such that
$k$ is on the non-physical plane for $\sqrt{n_k^2-k}$.
Note that (see Figure 1) 
$$k\in \tilde{S} \Rightarrow \Lambda (k)=\left \{ 
\begin{array}{cc}
\{ \ \} & \mbox{or}\\
\{ 0^2,1^2,\ldots ,n_k^2 \} & \mbox{or}\\
\{ n_k^2 \} .&
\end{array} \right .
$$

\begin{figure}[ht]
\vspace{3.0in}
\noindent
\caption{
Possible values for $\Lambda (k)$.
}
\label{Fone}

\vspace{-0.8in }

\vspace{0.6in }

\end{figure}
Denote $\Lambda (k)^c$ to be
the complement of $\Lambda (k)$ within the integers. 
We note for future reference the following formula:
\begin{equation}
\Re \sqrt{n^2-k}=
\pm \frac{((\Im k)^2 +(n^2-\Re (k))^2)^{1/4}}{2}
(1-\frac{1}{\sqrt{1+(\Im k/(n^2-\Re k))^2}}),\label{re}
\end{equation}
with the plus (resp. minus)
sign appearing when $n\in \Lambda (k)^c$
(resp. $n\in \Lambda (k)$). 

\begin{lemma}\label{Lone}
Let $\psi_1,\psi_2$ be smooth functions of bounded support on $\Omega$
that depend on $x$ alone, and with values in $[0,1]$.
Then the mapping from ${\bf C}-[0,\infty ) $ to $\bo$ given by
$$k\ra \psi_2 R_0(k)\psi_1 $$
 extends to a meromorphic function on $S$.
Also, the same is true for the mappings
$$k\ra \frac{\pa }{\pa x}\psi_2 R_0(k)\psi_1 ,\
k\ra \frac{\pa }{\pa y}\psi_2 R_0(k)\psi_1
.$$
Furthermore, for
$k\in \tilde{S}_{\al}$ on the physical sheet,
\begin{equation}
\| \psi_2R_0(k)\psi_1
\|_{L^2\ra L^2} \leq C\ln(|k|)(1+|k|)^{i/2},\ i=0,2,\label{i}
\end{equation}
where $C$ depends only on  $\al ,\psi_1,\psi_2$, and 
\begin{equation}
\| \partial^i/\partial x^i\psi_2R_0(k)\psi_1
\|_{L^2\ra L^2} \leq C(1+|k|)^{i/2},\ i=1,3\label{j}
\end{equation}
where $C$ depends only on  $\al ,\psi_1,\psi_2$.
\end{lemma}
\Pf
The analytic extension of $\psi_2R_0(k)\psi_1 $
and it's first partial derivatives follows
immediately from the  compactness of the support
of $\psi_1G_k\psi_2$, together with the pointwise
meromorphiciticity in $k$ of $G_k$.

To prove Eqs.~\ref{i},~\ref{j},
 fix $k$  on the physical plane. If $|\Im k|>1$, then
Eqs.~\ref{i},~\ref{j} hold be the Spectral Theorem and interpolation.
Thus in what follows, we assume $|\Im k|\leq 1$.

Let P be the orthogonal projection of $L^2(\Omega_0)$ onto
the closed subspace spanned  by $\{ f(x)\cos (ny); f\in L^2(\bR ),
0\leq n\leq n_k\}$. Thus $(I-P)\Delta_0(I-P)$
is
a self-adjoint operator whose spectrum is $[(n_k+1)^2,\infty )$.
Since $k$ is on the physical branch of all of the square-root
functions associated to the Schwartz kernel of $(I-P)\Delta_0(I-P)$,
it follows that
 $((I-P)\Delta_0(I-P)-k)^{-1}$ is a bounded operator
on $L^2(\Omega_0)$ and the following estimate holds by the
Spectral Theorem:
\begin{eqnarray}
\| (I-P)R_0(k)(I-P)\|
& = &
\| ((I-P)\Delta_0(I-P)-k)^{-1}\|  \nonumber
  \\
& \leq & |k-(n_k+1)^2|^{-1}\nonumber\\
& \leq & \al^{-1}.\nonumber
\end{eqnarray}
The last  inequality follows
from the assumption that $|k-n^2|>\al$ for all $n$.
Hence
\begin{equation}
\| \psi_1(I-P)R_0(k)(I-P)\psi_2\|\leq \al^{-1}.\label{1}
\end{equation}

We now estimate
the  norm of the operator $\psi_1PR_0(k)P\psi_2$, whose Schwartz kernel
is
$$\psi_1 (x)\psi_2 (x')
\frac{1}{\pi \sqrt{-k}}e^{-\sqrt{-k}|x-x'|}+\psi_1 (x)\psi_2 (x')
\sum _{n=1}^{n_k}
\frac{1}{\sqrt{n^2-k}}
e^{-\sqrt{n^2-k}|x-x'|}\cos (ny) \cos (ny').$$

In what follows, $C$ will
denote various positive constants that are independant of $k,n$.
Then, since $|\Im k|<1$,
\begin{eqnarray}
\| \psi_1PR_0(k)P\psi_2\| & \leq &\frac{1}{\pi \sqrt{|k|}}+
\sum _{n=1}^{n_k}
\frac{1}{|\sqrt{n^2-k}|}
\nonumber\\
& \leq & \sum _{n=0}^{n_k}
\frac{1}{|\sqrt{n^2-k}|}
\nonumber\\
& \leq & C\ln (\Re k ).\label{2}
\end{eqnarray}
Since $P$ commutes with $R_0(k)$, it follows
that
$$\psi_1(I-P)R_0(k)(I-P)\psi_2+\psi_1PR_0(k)P\psi_2
=\psi_1R_0(k)\psi_2.$$
Hence, by combining Eqs.~\ref{1},~\ref{2} we obtain that
 as $k\ra \infty$ with $k\in \tR_{\al}$, we have
$$\| \psi_1R_0(k)\psi_2\| \leq C \ln (| k|).$$
Thus
Eq.~\ref{i} has been proven for $i=0$.
 The proof of Eq.~\ref{j}, with $i=1$, is similar.
For $i=2$, we write
\begin{eqnarray*}
 \frac{\partial^2}{\partial x^2}\psi_2R_0(k)\psi_1
 &=&  \frac{\partial^2}{\partial x^2}(\Delta_0+1)^{-1}[\Delta_0,\psi_2]
R_0(k)\psi_1\\
& + & (k+1)\frac{
\partial^2}{\partial x^2}(\Delta_0+1)^{-1}
\psi_2R_0(k)\psi_1\\
& +&\frac{
\partial^2}{\partial x^2}
(\Delta_0+1)^{-1}
\psi_2\psi_1.
\end{eqnarray*}
Then the desired estimate follows from the estimates for $i=0,1$.

The proof for $i=3$ is similar.

\

We now prove the existence of a meromorphic extension of
$R(k)$. The argument follows  closely along the lines of
the corresponding result for exterior domains found in
 \cite{SjZ}. For a proof of this result for more
general perturbations of the cylinder, see \cite{Mel}.
\begin{prop}
Let $\chi \in C_0^{\infty}(\Omega )$.
 Then
the mapping from ${\bf C}-[0,\infty ) $ to $\bo$ given by
$$k\ra \chi (\Delta -k)^{-1}\chi $$
extends to a meromorphic function in $S$. At each pole $k_0$,
the coefficients of the negative powers of $(k-k_0)$ in
 the Laurent series are finite rank operators. 
\end{prop}
\Pf
We define an approximation of $R(k)$ as
follows. Assume for the moment that $k$ is on the physical
branch of $S$.
Let
\begin{equation}
R_{a}(k)=\tau_1 R(k_0)\chi_1 +\tau_2 R_0(k)\chi_2.\label{approx}
\end{equation}
Here $k_0$ is a parameter to be chosen below.

We have
\begin{equation}
(\Delta -k)R_a(k)= I+K,\label{mero}
\end{equation}
with
\begin{equation}
K=(k_0-k)\tau_1R(k_0)\chi_1+[\Delta,\tau_1]R(k_0)\chi_1
+[\Delta,\tau_2]R_0(k)\chi_2.\label{K}
\end{equation}
By Eq.~\ref{mero}  we have for
$k\in \bC-[0,\infty )$:
$$R_a=(\Delta-k)^{-1}(I+K).$$
By Eqs.~\ref{K},~\ref{rho} we have $\rho K=K$, hence
$$R_a\rho =(\Delta-k)^{-1}\rho (I+K\rho ).$$
For $k=k_0$ and
$ \ Im(k_0)>>0$, we have by the Spectral Theorem that $\| K\rho \|_{L^2\ra L^2}<1$
and hence we can write
\begin{equation}
\rho R_a(k)\rho (I+K\rho )^{-1}=\rho R(k)\rho .\label{mero1}
\end{equation}
Fix such a $k_0$.

Next we observe from Eq.~\ref{approx} that
$k\ra \rho R_a(k)\rho $ extends meromorphically to $S-\{ 0^2,1^2,\ldots \}$, 
with values in $\bo$. For the terms involving
$R_0(k)$, this follows from Lemma~\ref{Lone}, while for the term
involving $R(k_0)$,  note that the function
$k\ra (k_0-k)$ extends to the function $k\ra (k_0-\Pi k)$,
which is analytic on $S$. It follows
that the
 meromorphy of $\rho R(k)\rho $ is
equivalent to meromorphy of
$(I+K\rho)^{-1}$.

On the other hand, since $\chi_1$ and
$\rho$ are compactly supported, it follows
that $K\rho $ is an analytic compact operator-valued function of $k$ on
$S-\{ 0^2,1^2,\ldots \}$.
Thus $\rho R(k)\rho $ is a finitely-meromorphic Fredholm family in $k$
\cite{Vain}, and Meromorphic Fredholm Theory thus implies $\rho R(k)\rho$
is meromorphic for
 $k\in S-\{ 0^2,1^2,\ldots \}$.

To prove $\rho R(k)\rho$ is meromorphic in a neighbourhood of  
the threshold $L^2$, $L\in {\bf Z}$, 
one applies the argument above to the function $z\ra \rho R(L^2-z^2)\rho$
in a neighbourhood of $z=0$.

Finally, it is easy to see that
the function $\rho$ can be replaced by any smooth cutoff function.
This completes the proof. $\Box $\hfill

\

Next, we note
the following result due to Melrose
(\cite{Mel}, Prop. 6.28).
 As $z\ra 0$ for integer $L$,
\begin{equation}
\rho (\Delta -L^2-z^2)^{-1}\rho =\frac{A}{z^2}+\frac{B}{z}+C,\label{melr}
\end{equation}
where $A$ is the orthogonal projection onto the $L^2(\Omega )$ eigenspace
associated to the energy level $L^2$, 
 $B$ is a projection operator related to the generalised 
(non-$L^2(\Omega )$)
eigenfunctions associated to the energy level $L^2$, and $C$ is an operator
bounded near $z=0$.
We will use this asymptotic formula to provide upper bounds on the resolvent
in Lemma~\ref{Lfive}.

We now define the resonances of $\Delta$ to be the
poles $\rho R(k)\rho$ in $S-\{ L^2,\ L\in {\bf Z}\}$, in union with
any $L^2$ eigenvalues occuring at the thresholds. 
We define the multiplicity of a non-threshold 
resonance $k_j$ as the rank of the projection 
$$\int_{\gamma}\rho R(k)\rho dk$$
for a sufficiently small contour $\gamma $ about $k_j$. The multiplicity of
the any resonance occurring at a threshold is defined to be the dimension of
the corresponding eigenspace.

\end{section}
\begin{section}{Estimates on Fredholm Determinant} 

In what follows, let $C$ be various positive constants.
Let $K$ be as in the proof of Prop. 1.
A simple  argument shows that $(I+K\rho )$ is
invertible if and only if $(I+(K\rho )^{3})$ is invertible.
On the other hand,
since $K\rho$ is a pseudodifferential operator
of order -1 in $\bo$,
with compactly supported Schwartz kernel,
it follows that $(K\rho)^{3}$ is
trace class. Thus the Fredholm determinant
$det(I+(K\rho)^{3})$ is entire on $S$,
 and furthermore:

\begin{lemma}\label{Ltwo}
The non-threshold resonances of $\Delta$ (counted with their multiplicities)
are among the zeros of the  function
$$k\ra h(k)\equiv det(I+(K\rho )^{3}(k)),$$
counted with their multiplicities. 
\end{lemma}

The reader is referred to \cite{SZ} for a proof of this result.

\

The rest of this section is used to prove:
\begin{prop}
For $k\in \tilde{S}$, we have the estimate
\begin{equation}
| h(k)|\leq \frac{C\exp C|k|^{3/2}}{dist(k,\{ L^2, \ L\in {\bf Z}\} )^3}, 
\label{exp}
\end{equation}
with $C$ a positive constant independant of $k$.
\end{prop}
\Pf
Assume in what follows that $k\in \tilde{S}$, $k\neq L^2$ for $L\in {\bf Z}$.
We  apply the theory of characteristic values developed in
\cite{GK}, and adapted to exterior problems in \cite{M,Z2,V}.
The characteristic values $\mu_j(A)$ of a compact operator $A$ are
the eigenvalues, listed in decreasing order and counting
multiplicities, of the operator $|A|$. We recall the
following inequalities from \cite{GK}: $\mu_{j+k-1}(AB)\leq \mu_j(A)
\mu_k(B)$, $\mu_{j+k-1}(A+B)\leq \mu_j(A)+\mu_k(B),
\ \mu_j(AB)\leq \| A\| \mu_j(B)$.

We write $K\rho =K_1+K_2$, with
$K_2=[\Delta ,\tau_2]R_0(k)\chi_2\rho .$

Applying inequalities on Fredholm determinants appearing
in \cite{GK}, we get
\begin{eqnarray}
|det (I+(K\rho)^{3})|
\leq
&&
\!\!\!\!
\!\!\!\!
\!\!
det (I+4|K_1|^{3})^{6}
det (I+4|K_2|^{3})^{6}\nonumber
\\
\leq
&&
\!\!\!\!
\!\!\!\!
\!\!
(\prod_{j=1}^{\infty}(1+4\mu_j(|K_1|)^{3}))^{6}
(\prod_{j=1}^{\infty}(1+4\mu_j(|K_2|)^{3}))^{6}.
\label{det}
\end{eqnarray}
We shall estimate the terms on the right hand side of the last equation
with a  series of lemmas.
We estimate first the term involving $K_1$.
Recall that for $k\in S$,
\begin{equation}
K_1=(k_0-\Pi k)
\tau_1R(k_0)\chi_1\rho +
[\Delta,\tau_1]R(k_0)\chi_1\rho .\label{k1}
\end{equation}
\begin{lemma}\label{e1}
$$\prod_{j=1}^{\infty}(1+4\mu_j(|K_1|)^{3}) \leq Ce^{C|k|\ln |k|},\ 
\forall k \in S.$$
\end{lemma}
\Pf 
The argument here follows \cite{Z2}.
Since $\tau_1,\chi_1$ are compactly supported, it follows by standard
eigenvalue asymptotics for pseudodifferential operators \cite{shu}
that
$$\mu_j(|\tau_1R(k_0)\chi_1\rho|)\sim Cj^{-1},$$
 and
$$\mu_j(|[\Delta,\tau_1]R(k_0)\chi_1\rho |)\sim Cj^{-1/2}.$$
It follows that, denoting the largest integer below $x$ by $\lfloor x \rfloor$,
$$\mu_{j-1} (|K_1|)\leq C|k|\lfloor j/2\rfloor
^{-1}+C\lfloor j/2\rfloor^{-1/2}.$$
Hence
we get:
\begin{eqnarray*}
\mu_j(|K_1|^{3}) & \leq & (\mu_{(\lfloor j/3\rfloor+1) }
(|K_1|))^{3}\\
& \leq & (C|k|(\lfloor j/6+2\rfloor^{-1}
+C(\lfloor j/6\rfloor +2)^{-1/2})^{3}\\
& \leq & C|k|^{3}j^{-3}+Cj^{-3/2}.
\end{eqnarray*}
Note that $|k|^2<j$ is equivalent to $|k|^3j^{-3}<j^{-3/2}$.
Thus,
\begin{eqnarray*}
\prod_{j=1}^{\infty}(1+4\mu_j(|K_1|)^{3})
\ \ \leq &&
\\
\prod_{j\leq |k|^{2}}
(1+C|k/j|^{3}
&&
\!\!\!\!
\!\!\!\!
\!\!
)
\prod_{j> |k|^{2}}
(1+Cj^{-3/2})
\end{eqnarray*}
These two factors are bounded as in \cite{Z2}; we sketch the
argument. The first factor is bounded by comparing
it to
\begin{equation}
\exp \left(   \int_{1}^{|k|^2}\ln (1+C|k/x|^{3})dx \right) .\label{prod}
\end{equation}
Note that
\begin{eqnarray*}
\int_{1}^{|k|^2}\ln (1+C|k/x|^{3})dx & \leq & 
\int_{1}^{C|k|}\ln (1+C|k/x|^{3})dx + 2\int_{C|k|}^{|k|^2}
C|k/x|^3dx\\
& \leq & C|k|\ln (|k|).
\end{eqnarray*}
Thus Eq.~\ref{prod} is bounded by $\exp (C|k|\ln |k|)$.
The second factor is treated similarly.
Thus
\begin{equation}
\prod_{j=1}^{\infty}(1+4\mu_j(|K_1|)^{3})\leq e^{C|k|\ln |k|}.
\label{k12}
\end{equation}
Note that this estimate holds for all $k\in S$.

\

Next, we estimate the terms involving $K_2$ away from the thresholds.
\begin{lemma}\label{Lthree}
 Suppose $k\in \tR_{\al}$. Then
$\prod_{j=1}^{\infty}(1+4\mu_j(|K_2|)^{3})
<e^{C|k|^{3/2}}$,
 where  $C$ is
 some positive constant.
\end{lemma}

\Pf
The proof is an adaptation of the ``good half plane-
bad half plane'' argument found in \cite{Z2,V}. 

First, we assume $k\in \tilde{S}_{\al}$ is on the physical sheet, 
and assume without loss of generality that $|k|$ is large.
We have, by Lemma ~\ref{Lone},
\begin{eqnarray*}
\mu_j(K_2)& = & \mu_j(\rho (I+\Delta)^{-1}(I+\Delta )K_2)\\
& \leq & \mu_j(\rho (I+\Delta)^{-1})\| (I+\Delta )K_2)\| \\
&\leq & Cj^{-1}|k|^{3/2}.
\end{eqnarray*}
Now the  arguments leading to Eq.~\ref{k12} are easily adapted
to this case. In fact,
\begin{eqnarray*}
\prod_{j=1}^{\infty}(1+4\mu_j(|K_2|)^{3})
& \leq & \prod_{j=1}^{\infty}(1+C|k|^{9/2}/j^{3}).
\end{eqnarray*}

This last product is estimated as follows. First, one proves
\begin{equation}
\exp \left(   \int_{1}^{|k|^{3/2}}\ln (1+C|k|^{3/2}/x)^{3})dx
\right) \leq e^{C|k|^{3/2}}\label{int}
\end{equation}
as follows.
$$
\int_{1}^{|k|^{3/2}}\ln (1+C|k|^{9/2}/x^{3})dx
 =  \left ( 
\int_{1}^{|k|^{3/2}/\ln |k|}
+\int_{|k|^{3/2}/\ln |k|}^{|k|^{3/2}/10}
+\int_{|k|^{3/2}/10}^{|k|^{3/2}}\right )
\ln (1+C|k|^{9/2}/x^{3})dx.$$
The first and third integrals on the right hand side are easily shown
to be $O(|k|^{3/2})$, and the second integral is estimated as follows:
\begin{eqnarray*}
\int_{|k|^{3/2}/\ln |k|}^{|k|^{3/2}/10}
\ln (1+C|k|^{9/2}/x^{3})dx
& \sim &\int_{|k|^{3/2}/\ln |k|}^{|k|^{3/2}/10}
\ln (C|k|^{9/2}/x^{3})dx\\
& \leq & C|k|^{3/2}.
\end{eqnarray*}

Also it is easy to show that
\begin{equation}
\exp \left(   \int_{|k|^{3/2}}^{\infty}\ln (1+C|k^{3/2}/x|^{3})dx
\right) \leq e^{C|k|^{3/2}}.\label{int2}
\end{equation}
By Eqs.~\ref{int},~\ref{int2}, the lemma holds for $k$ on the physical
sheet.

Now suppose $k\in \tilde{S}_{\al}$ 
is on the non-physical sheet. There are two possible cases:
$\Lambda (k)=\{ 0,1, \ldots ,n^2_k\} $ or 
$\Lambda (k)=\{ n^2_k\} $. Suppose for now the first case.
We write
\begin{eqnarray*}
G_k(x,x',y,y')& = &
\frac{1}{\pi \sqrt{-k}}(e^{-\sqrt{-k}|x-x'|}+e^{\sqrt{-k}|x-x'|})\\
&+&\sum _{n=1}^{n_k}
\frac{1}{\sqrt{n^2-k}}
(e^{-\sqrt{n^2-k}|x-x'|}+e^{\sqrt{n^2-k}|x-x'|})\cos (ny) \cos (ny')\\
& -& 
\frac{1}{\pi \sqrt{-k}}e^{\sqrt{-k}|x-x'|}
-\sum _{n=1}^{n_k}
\frac{1}{\sqrt{n^2-k}}
e^{\sqrt{n^2-k}|x-x'|}\cos (ny) \cos (ny')\\
& + & \sum _{n=n_k+1}^{\infty}
\frac{1}{\sqrt{n^2-k}}
e^{-\sqrt{n^2-k}|x-x'|}\cos (ny) \cos (ny').
\end{eqnarray*}
Note first that
$$
e^{-\sqrt{n^2-k}|x-x'|}+e^{\sqrt{n^2-k}|x-x'|}=
e^{-\sqrt{n^2-k}(x-x')}+e^{\sqrt{n^2-k}(x-x')},
$$
and hence the operator $A_1$ whose Schwartz kernel is
$$
\frac{1}{\pi \sqrt{-k}}(e^{-\sqrt{-k}|x-x'|}+e^{\sqrt{-k}|x-x'|})
+\sum _{n=1}^{n_k}
\frac{1}{\sqrt{n^2-k}}
(e^{-\sqrt{n^2-k}|x-x'|}+e^{\sqrt{n^2-k}|x-x'|})\cos (ny) \cos (ny')
$$
will have rank $2n_k+2$. Thus the operator $[\Delta ,\tau_2]A_1\chi_2\rho$
will also have rank $2n_k+2$.
It follows now from Eq.~\ref{re} that for
$k\in \tilde{S}_{\al}$, 
$$\mu_j([\Delta ,\tau_2]A_1\chi_2\rho)\leq \left \{  
\begin{array}{cc}
e^{C|k|^{1/2}} & j\leq 2n_{k}+2\\
0 & j>2n_{k}+2.
\end{array}
\right .
$$ 
Also, observe that the operator $A_2$, whose Schwartz kernel is given by
\begin{eqnarray*}
-\frac{1}{\pi \sqrt{-k}}e^{\sqrt{-k}|x-x'|}&
-&\sum _{n=1}^{n_k}
\frac{1}{\sqrt{n^2-k}}
e^{\sqrt{n^2-k}|x-x'|}\cos (ny) \cos (ny')\\
& +&  \sum _{n=n_k+1}^{\infty}
\frac{1}{\sqrt{n^2-k}}
e^{-\sqrt{n^2-k}|x-x'|}\cos (ny) \cos (ny'),
\end{eqnarray*}
equals $R_0(\Pi k)$, ie. $R_0$ evaluated on the physical sheet.
Hence
\begin{eqnarray}
det(I+4|K_2(k)|^{3}) &\leq
&
det(I+16|[\Delta ,\tau_2]A_1\chi_2\rho |^{3}
)^{6} 
det(I+16|[\Delta ,\tau_2]A_2\chi_2\rho |^{3})^{6}\nonumber \\
& \leq & \prod_{j=1}^{2n_k+2}(1+e^{C|k|^{1/2}})\ \cdot \ e^{C|k|^{3/2}}\\
& \leq & e^{C|k|^{3/2}}.
\end{eqnarray} 
The last inequality holds because $n_k |k|^{1/2}\leq C|k|$.

For the case where
$\Lambda (k)=\{ n^2_k\} $, we write
\begin{eqnarray*}
G_k(x,x',y,y')& = &
\frac{1}{\pi \sqrt{-k}}e^{-\sqrt{-k}|x-x'|}+\sum _{n=1}^{n_k-1}
\frac{1}{\sqrt{n^2-k}}
e^{-\sqrt{n^2-k}|x-x'|}\cos (ny) \cos (ny')\\
&+&\sum _{n=n_k+1}^{\infty}
\frac{1}{\sqrt{n^2-k}}
e^{-\sqrt{n^2-k}|x-x'|}\cos (ny) \cos (ny')\\
& +& 
\frac{1}{\sqrt{n_k^2-k}}
(e^{-\sqrt{n_k^2-k}|x-x'|}-e^{\sqrt{n_k^2-k}|x-x'|})
\cos (ny) \cos (ny').
\end{eqnarray*}
The argument in this case is similar to the one for the case previous.
The details are left to the reader.

\

Proposition 2, for $k$ away from the thresholds,
now follows from Lemmas ~\ref{e1},~\ref{Lthree} and Eq.~\ref{det}.
We now prove bounds on the determinant near the thresholds. 
\begin{lemma}\label{thresh-b}
Let $L$ be any integer. Then for $k\in S$, $dist(k,L^2 ) \leq \al$, 
$$|det (I+(K\rho)^3)|\leq \frac{Ce^{C|k|^{3/2}}}{dist(L^2,k)^{3/2}},$$
with $C$ independant of $k,L$.
\end{lemma}
\Pf
The key observation is that the pole for $K$ at $L^2$ 
is simple with rank one residue.
In particular, note that
\begin{eqnarray*}
[\Delta ,\tau_2] \frac{1}{\sqrt{L^2-k}}e^{\sqrt{L^2-k}|x'-x|}&
=&\tau ''\frac{1}{\sqrt{L^2-k}}\\
& + &
\tau ''\frac{e^{\sqrt{L^2-k}|x'-x|}-1}{\sqrt{L^2-k}}\\
&+&2\tau 'sgn (x'-x)e^{\sqrt{L^2-k}|x'-x|},
\end{eqnarray*}
where $sgn (t)=1$ for $t>0$, $sgn (t)=-1$ for $t<0$.
Thus, we may write
$$K\rho =K_3+K_4,$$
where  
$K_3$ is a rank one operator with Schwartz kernel
\begin{equation}
\frac{1}{\sqrt{L^2-k}}\tau_2''(x)\rho (x')\chi_2 (x')\cos (ny)\cos (ny')
,\label{k3}
\end{equation}
and $K_4$ has pointwise bounded Schwartz 
kernel in a neighbourhood of the $L^2$.
In the argument that follows we use the inequality 
$$\mu_j (|A+B|^2)\leq 2\mu_j(|A|^2+|B|^2),$$
which follows from the quadratic form inequality $|A+B|^2\leq 2|A|^2+2|B|^2$
and a minimax argument. Fix $\ep \in (0,1/2)$.
\begin{eqnarray*}
\mu_j ((K\rho )^3) 
& = & \mu_j (|K\rho |^2)^{3/2}\\
& \leq & \mu_j (|K_3+K_4|^2 )^{3/2} \\
& \leq & 2^{3/2}\mu_j (|K_3|^2+|K_4|^2 )^{3/2} \\
& \leq & 2^{3/2}(\mu_{\lceil (1-\ep )j\rceil} (|K_3|)^2+
\mu_{\lfloor \ep j\rfloor+1} (|K_4|)^2 )^{3/2} .
\end{eqnarray*}
It follows from Eq.~\ref{k3} that
$$
\mu_{\lceil (1-\ep )j\rceil} (|K_3|)
=\left \{ 
\begin{array}{cc}
C/\sqrt{L^2-k}, & j=1\\
0, & j>1.
\end{array}
\right .
$$
Thus
\begin{eqnarray}
|det (I+(K\rho )^3)|& \leq & \prod_{j=1}^{\infty}(1+\mu_j((K\rho )^3))\nonumber\\
& \leq &\left (
1+2^{3/2}\left ( \frac{C}{(L^2-k)^{1/2}}+\mu_1(K_4)\right )^3
\right )\nonumber \\
& \cdot &\prod_{j=2}^{\infty}(1+
2^{3/2}\mu_{\lfloor \ep j\rfloor +1 }(K_4)^3).\label{prod2}
\end{eqnarray}
Now we analyse $\mu_j (K_4)$. Let $P$ be the orthogonal projection onto
the orthogonal complement of the subspace of $L^2(\Omega_0)$:
$$\{ f(x)cos (Ly); \int _{-\infty}^{\infty}
|f(x)|^2 dx<\infty \} .$$
Then we have $K_4=K_1+K_2P+K_5$, where $K_1$ and $K_2$ are as in the proof
of Prop. 2, and 
where $K_5$ has Schwartz kernel
\begin{eqnarray}
\rho (x)& \left ( \right .
\tau ''(x) &\left .  \frac{e^{\sqrt{L^2-k}|x'-x|}-1}{\sqrt{L^2-k}}
+2\tau 'sgn(x'-x)e^{\sqrt{L^2-k}|x'-x|}\right ) \nonumber \\
& \cdot &
\chi_2(x')\rho (x')
\cos (Ly)\cos (Ly').\label{HS}
\end{eqnarray}
Thus
\begin{eqnarray*}
\mu_n(K_4)^3& \leq &(\mu_{\lfloor n/3\rfloor +2}(K_1)
+\mu_{\lfloor n/3\rfloor}(K_2P)
+\mu_{\lceil n/3\rceil}(K_5))^3\\
& \leq & C((\mu_{\lfloor n/3\rfloor+2}(K_1)^3
+\mu_{\lfloor n/3\rfloor}(K_2P)^3
+\mu_{\lceil n/3\rceil}(K_5)^3).
\end{eqnarray*}
Hence, by applying inequalities for Fredholm determinants as in Eq.~\ref{det},
\begin{eqnarray*}
\prod_{j=1}^{\infty}(1+
2^{3/2}\mu_{\lfloor \ep j-1\rfloor }(K_4)^3)
&\leq &
(\prod_{j=1}^{\infty}(1+
C\mu_{\lfloor \ep j/3\rfloor +1}(K_1)^3))^{6}\\
&\cdot &(\prod_{j=1}^{\infty}(1+
C\mu_{\lfloor \ep j/3\rfloor }(K_2P)^3))^{36}\\
&\cdot &
(\prod_{j=1}^{\infty}(1+
C\mu_{\lfloor \ep j/3\rfloor }(K_5)^3))^{36}.
\end{eqnarray*}
The first of these products is estimated 
exactly as in the proof of Lemma ~\ref{e1}, 
and the estimate for the second of these products is derived similarly
to the estimate for $det (I+4|K_2|^3)$ away from the thresholds (noting
that $\mu_n(K_2P)\leq \mu_n (K_2)$).
Finally,
\begin{eqnarray*}
\prod_{j=1}^{\infty}(1+
C\mu_{ j }(K_5))^3
& = & \exp (\sum (\ln (1+\mu_j ((K_5)^3))))\\
& \sim & \exp (\sum \mu_j ((K_5)^3))\\
& \leq & \exp ((\sum \mu_j (|K_5|^2)^2)^{1/2}\cdot
(\sum \mu_j ((K_5)^2))^{1/2})\\
& \leq & C.
\end{eqnarray*}
The last inequality holds because,
by Eq.~\ref{HS}, the Hilbert-Schmidt norms  (see, eg., \cite{RS} Vol. 1)
of $K_5$ and 
$K_5^2$ are  bounded  by a bound independant of $k$, for $k$ in a small
neighbourhood of $L^2$.
Thus
$$
\prod_{j=1}^{\infty}(1+
2^{3/2}\mu_{\lfloor \ep j-1\rfloor }(|K_4|)^3)\leq Ce^{C|k|^{3/2}}.$$
Proposition 2 now follows from Lemmas
~\ref{e1},~\ref{Lthree},~\ref{thresh-b}.
\end{section}
\begin{section}{Resolvent estimate and bounds on number of resonances}

Proving both Theorems 1 and 2 requires lower bounds on the Fredholm determinant
studied in the previous section. For this
we use the following lemma, which is an adaptation of
 an argument found in Cartwright (\cite{cart}, p.89-91). The Cartwright
result has previously been used in the scattering theoretic context by 
\cite{PZ}.
\begin{lemma}\label{complex} Let $k=re^{i\theta}$.
Suppose the function $g$ is analytic in the sector
$\{ \theta \in (0,\pi  )\} $, and 
satisfies 
$$|g(k)|<C\exp (C|k|^{3/2}).$$
Let $\phi: {\bf R}\ra {\bf R}$ 
be any increasing, real valued function such that $\lim_{x\ra \infty}\phi (x)
=\infty$.
Then, for any $M>0$,  there exists $R_2=R_2(M)$ such that for each 
$r>R_2$,
$$|g(k)|>\exp (-|k|^{5/2}\phi (|k|)),$$
except perhaps in a set of $\theta $, denoted $\theta_r$, with 
$|\theta_r|<1/(Mr^{1/2})$. Here $|A|$ denotes the Lebesgue measure of $A$.
\end{lemma}
\Pf 
By Carleman's Formula \cite{L},
we have 
\begin{eqnarray*}
\sum_{1\leq r_n\leq R}\frac{\sin \theta_n}{r_n}(1-\frac{r_n^2}{R^2})& =
& \frac{1}{\pi R}\int_{0}^{\pi }\ln |g(Re^{i\theta })|\sin \theta
d\theta \\
& +& \frac{1}{2\pi}\int_1^R (\ln |g(y)|+\ln |g(-y)|)(\frac{1}{y^2}-\frac
{1}{R^2})dy+\chi (R),\\
\end{eqnarray*}
where $\chi (R)=O(1)$ as $R\ra \infty$, and $r_ne^{i\theta_n}$ are the
zeros of $g(z)$ in upper half space.
Set $\ln_+(x)=\max (\ln (x),0)$.
Then it follows that
\begin{eqnarray}
\frac{1}{\pi R}\int_{0}^{\pi }\ln |g(Re^{i\theta })|\sin \theta d\theta
& 
\geq& -\frac{1}{2\pi}\int_1^R (\ln_+ |g(y)|+\ln_+ g|(-y)|)
\frac{1}{y^2}dy-O(1)
\nonumber \\
&\geq & -C\int_1^R y^{3/2}\frac{1}{y^2}dy-O(1)\nonumber \\
&\geq & -C R^{1/2}-O(1)\nonumber \\
&\geq & -C R^{1/2} ,\ R>R_0,
\label{one} 
\end{eqnarray}
for some constant $R_0$.
Now fix $M>0$ and 
suppose now that for each $r$, 
there exists a set $\theta_r$ of measure
at least $1/Mr^{1/2}$ such that for $\theta \in \theta_r$,
$$|g(k)|<\exp (-|k|^{5/2}\phi (|k|)).$$
Then we have
\begin{eqnarray}
\frac{1}{\pi R}\int_{0}^{\pi }\ln |g(Re^{i\theta })|\sin \theta d\theta
& \leq & CR^{1/2}-2R^{3/2}\phi (R)
\int_{0}^{1 /(2MR^{1/2})
 }\sin \theta d\theta \nonumber \\
& \leq & CR^{1/2}-2R^{3/2}\phi (R)
(1-\cos (\frac{1}{2MR^{1/2}}))\nonumber \\
& \leq & CR^{1/2}-R^{1/2}\phi (R) /(4M^2) .
\label{two}
\end{eqnarray}
Comparing Eqs.~\ref{one},~\ref{two}, we derive a contradiction for 
$R>R_1$ for some $R_1=R_1(M)$. Setting $R_2=\max (R_0,R_1)$, we obtain
for $|k|>R_2$,
$$|g(k)|>\exp (|k|^{5/2}\phi (|k|)),$$
except for $\arg (k)\in \theta_{|k|}$, with $|\theta_{|k|}|<1 /(M|k|^{1/2})$.

The lemma is proven.

\

{\em Proof of Theorem 2:}

We can, without loss of generality, assume $|k|>R_2$,
where $R_2$ will be determined by Lemma~\ref{complex}.
The proof will apply Jensen's formula for meromorphic functions
(\cite{cart}, p.9) to the function 
$$h(k)=det(I+(K\rho )^3)$$
to obtain upper bounds on the number of resonances in a set of disks
that form a cover for the following subset of $\tilde{S}$:
$$\{ k\in S: dist(k,\mbox{physical plane})<1+\frac{1}{2}\sqrt{|k|}, \ 
|k|>R_2\}.$$

We consider separately the two cases:

A) the part of $\tilde{S}$  which is a continuation
from the upper half of the physical plane,

B) the part of $\tilde{S}$  which is a continuation
from the lower half of the physical plane.

We treat case A; the argument for case B is similar.
 To obtain the necessary
lower bounds on $h(k)$, we first 
apply Lemma~\ref{complex} and  Proposition 2 to a sector that is slightly
shifted away from the positive real axis. Setting $M=100$ and $\phi(x)=\ln x$ 
in 
Lemma~\ref{complex}, we get  for $|k|>R_2$ 
$$|h(k)|>\exp (-|k|^{5/2}\ln|k|),$$
except for $\arg (k)\in \theta_{|k|}$, with $|\theta_{|k|}|<1/(100|k|^{1/2})$.
Fix a positive integer $L$ with $L^2\geq R_2 $. 
Thus 
setting $\Re (k_0)= L^2+L$,
 one can choose $\Im k_0 < L/99$ such that
\begin{equation}
|h(k_0)|>\exp (-|k_0|^{5/2}\ln |k_0|).\label{k0}
\end{equation}
We now apply the Minimodulus Theorem of Cartan 
to obtain a lower bound
on $|h(k)|$ near the real axis. The following version of the theorem
can be easily deduced from the arguments found in (see \cite{L}, p.21-22):
 if $g$ is analytic in $B(0,R)\subset {\bf C}$ and
$|g(0)|>0$, then for any $r<R$ one has
$$|g(z)|>|g(0)|^{1+H}\left ( \max_{|z|=R}|g(z)|\right )^{
-H},\ \ H=\frac{2r}{R-r}+\frac{\ln (3e/2\eta )}{\ln (R/r)} ,$$
this estimate valid in $B(0,r)$ outside an exceptional set
of disks whose summed
radii is less that $4\eta R$. 
We set $R=L$,  $r=L/(10)$,
$\eta =1/(4L)$. Applying the Cartan theorem and Prop. 2,
 there exists  $k_1\in B(L^2+L,1)\subset S$ such that
\begin{eqnarray}
|h(k_1)|& > &\exp (-C|k_1|^{5/2}\ln |k_0|\ln |k_1|)\nonumber \\
& > & \exp (-C|k_1|^{5/2+\ep}).
\label{k11}
\end{eqnarray}
Note that the disk $B(k_1,.9L)$ does not contain any thresholds.
We apply Prop. 2 and Jensen's formula
to $h(k)$ on the disk $B(k_1,.9L)$ to conclude that the number of zeros,
counting multiplicities, in the disk $B_L\equiv B(k_1,.8L)$ is bounded by
$C|k_1|^{5/2+\ep }$.

Next, we bound the number of zeros in a neighbourhood of the threshold
$L^2$. Suppose first that $h(k)$ has a pole at $k=L^2$. Then clearly
there exists $k_2\in \tR$, with $dist (k_2, L^2)<1$, such that 
$|h(k_2)|>1$.  The function $z\ra h(L^2-z^2)$
 is meromorphic in the disk
$\{ z: |z-\sqrt{L^2-k_2}|<\sqrt{.9(L+1)}\} $; here we view the disk as
lying in the  complex plane and $|*|$ is the standard absolute value function. 
The only pole for $z\ra h(L^2-z^2)$
 in this disk is at $z=0$, and by Prop. 2 the pole has order at most 3. Hence
by Jensen's formula,
the number of zeroes in the disk 
\begin{equation}
\{ z: |z-\sqrt{L^2-k_2}|<\sqrt{.8(L+1)}\} \label{disk}
\end{equation}
is bounded by $C|k_2|^{3/2}$. We label the disk in $S$ corresponding to
Eq.~\ref{disk} as $\tilde{B}_L$.

Next suppose that $h(k)$ has no pole at $L^2$. It follows that 
$z\ra h(L^2-z^2)$ is analytic in the disk $\{ z: |z|<\sqrt{L}\}$.
By Prop. 2 and Lemma ~\ref{complex}, there exists $z_1$ with $|z_1|
<\sqrt{L/100} $ such that $|h(z_1)|>e^{-|L^2|^{5/2}\ln |L^2|}$. Applying the 
Cartan theorem, we obtain $z_2$ such that $|z_2|<1$ and
$|h(z_2)|>e^{-C|L^2|^{5/2+\ep}}.$ Now applying Jensen's formula as above,
the number of zeros on the disk $\{ z:|z-z_2|<\sqrt{.8(L+1)}\}$ 
is bounded by $CL^{5/2+\ep}$. Again in this case we label the corresponding 
disk in $S$ as
$\tilde{B}_L$.

Theorem 2 now follows by noting
$$\{ k\in \tilde{S},\ dist (k,\mbox{physical sheet})< \frac{\sqrt{|k|}}{2}+1, \
R_2<|k|<r\}
\subset \bigcup_{L=\lfloor R_2\rfloor ^2}^{\lceil r\rceil ^2}(B_L\cup \tilde{B}_L).$$

\begin{lemma}
\label{Lfive}

{\bf A:} 
For any $t,\ep ,\al>0$,
 there exists a constant
$C$ dependant on $\ep, \al$ but independant of $t$ such that
\begin{equation}
\| \rho R(k)\rho \| \leq Ce^{Ct|k|^{5/2+\ep}},
\label{fu4}
\end{equation}
for all $
k\in \tR_{\al}-\cup_{k_j}B(k_j,|k_j|^{-t})$ with $|\Im (k)|<|k|^{1/2}/2$,
where $k_j$ are among the resonances of $\Delta$.

{\bf B:}
For $L$ any integer, we have the following estimates at the threshold
$L^2$ for any $\ep ,\delta >0$. 
	If there exist no resonances in the disk $B(L^2,\delta )$, then
\begin{equation}
\| \rho R(k)\rho \| \leq \frac{Ce^{C|k|^{5/2+\ep}\ln (1/\delta )}}
{|L^2-k|\delta^{7/2}}.\label{vogel}
\end{equation}
If the only resonances in the disk $B(L^2,\delta )$ are precisely at
 $k=L^2$, then
\begin{equation}
\| \rho R(k)\rho \| \leq \frac{Ce^{C|k|^{5/2+\ep}\ln (1/\delta )}}
{|L^2-k|^{2}\delta^{5/2}}.\label{bommel}
\end{equation}
Here $C$ is independant of $k,L$.
\end{lemma}
\Pf
As in \cite{Z3}, we bound the resolvent in terms of Fredholm determinants.
In  what follows, $C$ will denote various positive constants.
For simplicity we set $\al =1$.
We recall:
\begin{equation}
\rho R_a(k)\rho (I+K\rho )^{-1}=\rho  R(k)\rho .\label{mero2}
\end{equation}
We begin by estimating the resolvent away from the thresholds, so 
our analysis will be conducted on $\tilde{S}_{\al}$.

It follows from Eq.~\ref{approx} and Lemma~\ref{Lone} that for
 $k\in \tilde{S}_{\al}$,
\begin{equation}
\| \rho R_a\rho \|_{L^2\ra L^2}\leq {C|k|^{1/2}}.\label{b1}
\end{equation}
To bound $(1+K\rho )^{-1}$, we proceed as follows:
from \cite{GK}, Thm.5.1, Ch.5, we have
\begin{equation}
\| (I+K\rho )^{-1}\|_{L^2\ra L^2}
\leq |det(I+(K\rho )^{3})|^{-1}det(I+|K\rho |^{3})^{3}.\label{r1}
\end{equation}
By the proof of Proposition 2, we have
\begin{equation}
det(I+|K\rho |^{3})^{3}\leq e^{C|k|^{3/2}},\ k\in \tilde{S}_{\al}.\label{fu1}
\end{equation}

We now obtain a lower bound on $|h(k)|= |det(I+(K\rho )^{3})|$.
Set $M=\frac{1}{100}$ and fix $R_2=R_2(M)$ in Lemma~\ref{complex}.
In what follows, we assume without loss of generality that 
$|k|>R_2$. We will also prove the result 
 only for the portion
of the non-physical branch in $\tilde{S}$ that is reached by a path from the 
upper half space.

Arguing as in the proof of Theorem 2, for
$L$ any non-negative integer, we have
that there exists  $k_1\in B(L^2+L,1)$ such that
$
|h(k_1)|
 >  \exp (-|k_1|^{5/2}\ln |k_1|).
$
Applying the Cartan Theorem with $R=.9L$, $r=.8L$, and $\eta =(2L)^{-2t-1}/4$,
 we
get
\begin{equation}
|h(k)|
 >  \exp (-(2t+1)C|k|^{5/2+\ep}), \label{bouma}
\end{equation}
for $k$ in $B(k_1,.8L)$ but outside an exceptional set of disks of radius
no larger than $(2L)^{-2t}$. 
We decompose the system of disks into the union
$\cup U_j$, where $U_j$ are connected and mutually disjoint.
We can assume that each $U_j$ contains a resonance, which we label $k_j$.
For if not, then Eq.~\ref{fu4} holds on $U_j$ by
the Maximum Principle. Using the inequality $(2L)^{-2t}<|k_j|^{-t}$,
it then follows that for each $j$,
$U_j \subset B(k_j, |k_j|^{-t})$.

We now obtain lower bounds on $h$ in a
neighbourhood of the thresholds. 
Choose $k_1$ with $\Re k_1=L^2$ and $\Im k_1\in (1,\frac{L+1}{100})$, such that
$$|h(k_1)|> \exp (-|k_1|^{5/2}\ln |k_1|).$$
Suppose first that $h(k)$ has a pole of order $j$ at $L^2$, with $j=1,2$
or 3. Applying the Cartan theorem to the function 
$z\ra z^jh(L^2-z^2)$, in the disk $\{ |z-\sqrt{L^2-k_1}|<R\}$
 with $R=1.2\sqrt{L}$, $r=\sqrt{L}$, and $\eta =1/(4R(2L)^{2t})$
we obtain
\begin{equation}
|h(k)|>\frac{e^{-Ct|k|^{5/2+\ep}}}{(L^2-k)^{j/2}},\label{kalou}
\end{equation}
for $k$ in $B(L^2,.9L)$ but
outside a union
of disks with summed radii no greater than
$(2L)^{-2t}$.
On the other hand, if $z\ra h(L^2-z^2)$ is regular at $z=0$, then 
 we can apply the Cartan theorem directly to obtain Eq.~\ref{kalou}
holding in $B(L^2,.9L)$.
In either case,
 the inequality appearing in Eq.~\ref{fu4} now
holds in  $\tilde{S}_1$
 in the complement
of the system of disks. Arguing as above, we can assume that the union of disks 
is of the form $\cup B(k_j, |k_j|^{-t})$.
Part A of the lemma has been proven.

We now prove Eq.~\ref{vogel}. Thus
 suppose there exists no 
$L^2(\Omega )$ eigenvalue in the disk $B(L^2,\delta )$. 
Using Prop.2 and Eq.~\ref{green} and using the Cartan theorem 
as above, we have
the following upper bound:
$$\| \rho (\Delta -L^2-z^2)^{-1}\rho \| <
\frac{Ce^{C|k|^{5/2+\ep}\ln (1/ \delta)}}{\delta ^4},\
 |z|\in (\sqrt{\delta /3},\sqrt{2\delta /3}).$$
Also, by
Eq.~\ref{melr}, the function $z\ra z\rho (\Delta -L^2-z^2)^{-1}\rho $
is analytic in $\{ z: |z|<\sqrt{2\delta /3}\} $. By the previous inequality
we have
$$\| z\rho (\Delta -L^2-z^2)^{-1}\rho \| <
\frac{Ce^{C|k|^{7/2+\ep}\ln (1/\delta )}}{\delta ^6}, \ |z|=\sqrt{2\delta /3}.$$
Hence Eq.~\ref{vogel} holds by the Maximum Principle.
The proof for Eq.~\ref{bommel}
is similar.
$\Box $\hfill

\begin{lemma}\label{8}
Let $p$ be an integer with $p>2$ and $q\in [0,p/2)$. Then there exists a positive constants
$M_{p,q},C_p$ such that if
$m>M_{p,q}$, and if
$f(k)$
is an analytic function  in a region $\Gamma$
in $\bc$, with
$$\Gamma\equiv \{ \Re(k)\in [m-2m^{-q},m+2m^{-q}],
-\frac{1}{(\Re k)^p}\leq \Im k\leq \frac{1}{(\Re k)^{2p}}
\} ,$$
and if $f$ satisfies the estimates:

A: $|f(k)|\leq e^{|k|^p},$

B: $|f(k)|\leq 1/\Im (k)$ for $\Im k>0$,

then for $k\in [m-m^{-q},m+m^{-q}]$,
we have
\begin{equation}
| f(k)| \leq C_p|k|^{2p}.\label{real}
\end{equation}
\end{lemma}
\Pf
In what follows, $C_p$ will denote various constants
independant of $m,k$, while $C$ will various constants independant of $m,k,p$.
We use  an argument based on the Maximum Principle. Below, we will construct
a family of functions $F_{\al}(k)$ parametrised by $\al$ such that

1: $F_{\al}
$ is analytic on $\Gamma$,

2: $|F_{\al}|<e$ on $\Gamma$,

3: on the interval $[m-m^{-q},m+m^{-q}]$, $|F_{\al}|>1/2$,

4: on $\{z\in \Gamma:|z-m|\geq \frac{3}{2}m^{-q}\} $, we have
$|F_{\al}|\leq  C|k|^p\exp (-C|k|^{2p-2q})$.

Assuming such $F_{\al}$ exist, consider the function on $\Gamma$:
$$h(k)\equiv f(k)F_{\al}(k)\exp (-ik^{2p+1}).$$
On the curve $\Im k=- {1}/{(\Re k)^p}$,
there exists a positive constant $M_1$ such that $\Re k>M_1$
implies $\Im (k^{2p+1})<-(\Re k)^p$ and $|k|^p-(\Re k)^p <C$.
Thus
we have
\begin{eqnarray*}
|h(k)| & \leq & \exp (|k|^p)\cdot e\cdot  \exp (-(\Re k)^p)\\
 & \leq & C.
\end{eqnarray*}
On the curve $\Im k= {1}/{(\Re k)^{2p}}$,
we have for $\Re k>M_1$
\begin{eqnarray*}
|h(k)| & \leq & (\Re k)^{2p}\cdot e\cdot  |\exp (C_p)|\\
 & \leq & C_p(\Re k)^{2p}.
\end{eqnarray*}
On the curve $\Re k= m+2m^{-q}$,
we have
\begin{eqnarray*}
|h(k)| & \leq & \exp ((m+2m^{-q})^p)\cdot C|k|^p\exp (-C|k|^{2p-2q})
\cdot |\exp (C_p)|\\
& \leq & C_p|k|^p\exp(|k|^{p-2q}).
\end{eqnarray*}
Noting that $p>2q$, it follows that
 there exists $M_2 >0$ such that if $m>M_2$, then
$|h(k)|<C_p$.
Similarly, we have
on the curve $\Re k= m-2m^{-q}$,
we have (assuming $m>M_2$)
\begin{eqnarray*}
|h(k)|  & \leq & C_p.
\end{eqnarray*}
Choose $M_3$ such that $\Re k>M_3-2$ implies $C_p(\Re k)^{2p}>C$
with $C$ the maximum of the various $C$'s above. Setting $M_{p,q}=\max (M_1,M_2,M_3)$,
we have that for $m>M_{p,q}$,
it follows by the Maximum Principle that $|h(k)|\leq C_p(m+2m^{-q})^{2p}$ on
$\Gamma$. Since $|F_{\al}\exp(-ik^{2p+1})|>1/2$ on the interval
$[m-m^{-q},m+m^{-q}]$, Eq.~\ref{real} follows.

It remains to prove the existence of $F_{\al}$.
Let $\psi \in C_0^{\infty} (\bR)$ by defined so that
$\psi =1$ on $[m-1.1m^{-q},m+1.1m^{-q}]$, and $\psi =0$ on $(-\infty, m-1.2m^{-q}]\cup
[m+1.2m^{-q},\infty )$. Define
$$F_{\al}(z)=(\pi \al^{-2})^{-1/2}
\int_{{\bf R}}\exp (\frac{-(x-z)^2}{\al^2})\psi (x)dx.$$
The analyticity of $F_{\al}$ follows immediately. To prove
Property 2, note first that
\begin{equation}
(\pi \al^{-2})^{-1/2}
\int_{{\bf R}}\exp (\frac{-x^2}{\al^2})dx=1.
\end{equation}
Thus, setting $z=u+iv$, with $u,v\in \bR$, it is easy to see that
$$|F_{\al}(z)| \leq \exp |v^2/\al^2|.$$
Setting $\al =(m+2m^{-q})^{-p}$, Property 2 follows.

To prove Property 3, suppose $z\in [m-m^{-q},m+m^{-q}]$.
Thus
\begin{eqnarray*}
|F_{\al}(z)-1| & = & \pi^{-1/2}\int_{\bR}e^{-y^2}|\psi(\al y+z)-1|dy\\
& \leq & \pi^{-1/2}\int_{|y|>.1m^{-q}/\al}e^{-y^2}dy\\
& \leq & 1/2
\end{eqnarray*}
since $m^{-q}/\al$ is large.

For Property 4, assume $z=u+iv\in \Gamma \cap \{ \zeta: |\zeta-m|
>3/2m^{-q}\} .$
Then
\begin{eqnarray*}
|F_{\al}(z)| & \leq & (\pi \al^{-2})^{-1/2}
|\exp (v^2/\al^2)|
\int_{\bR}\exp (\frac{-(x-u)^2}{\al^2})\psi (x)dx\\
&\leq & (\pi \al^{-2})^{-1/2}
\cdot e\cdot \int_{[m-1.2m^{-q},m+1.2m^{-q}]}\exp (\frac{-(x-u)^2}{\al^2})dx\\
& \leq & C \al^{-1}\exp (-(.3m^{-q})^2\al^{-2})\\
&\leq & C |k|^{p}\exp (-C|k|^{2p-2q}).
\end{eqnarray*}
The proof of Lemma 5 is complete.

\vspace{0.5in}

\noindent Proof of Theorem 1:

In view of Lemmas~\ref{Lfive},~\ref{8}, we set $p=3$.
We can (increasing $M_{p,q}$ if necessary)
suppose $\| \rho R(k)\rho \| \leq e^{|k|^3}$
away from $\cup (k_j,|k_j|^{-t})$.
Let $q\in [0,p/2)$, and let $M_{p,q}$ be as in Lemma~\ref{Lfive}.
Assume the hypotheses of the theorem; hence
 the operator valued function $\rho R(k)\rho$ is analytic on
$$\{ k\in S: \ dist(k,[m-2m^{-q}
,m+2m^{-q}])< 2(m-2)^{-3}\} . $$
Setting $t=2p$ in Lemma~\ref{Lfive}, we obtain
$$\| \rho R(k)\rho \| \leq e^{|k|^{3}}$$
in the region
$${\cal G}\equiv \{ k\in S: \ dist(k,[m-2m^{-q}
,m+2m^{-q}])< (m-2)^{-3}\} . $$

Let $\tilde{\Gamma}\subset {\cal G}$ be an open subset
such that
the projection $\Pi$, restricted to $\tilde{\Gamma}$,
is an isometry 
onto $\Gamma$,
with $\Gamma$ as in Lemma~\ref{8}. Thus $\tilde{\Gamma}$ lies
on one of the branches of ${\cal G}$, and its intersection
with the physical plane
will be  non-empty and consist of one of the two sets
$$k: \Gamma_-\equiv
\{ \Re(k)\in [m-2m^{-q},m+2m^{-q}],
-\frac{1}{(\Re k)^3}\leq \Im k\leq 0
\} ,$$
or
$$k: \Gamma_+\equiv \{ \Re(k)\in [m-2m^{-q},m+2m^{-q}],
0\leq \Im k\leq \frac{1}{(\Re k)^{6}}
\} .$$
Assume for the moment that the intersection is $\Gamma_+$. Then
we have shown that estimate A of the previous lemma holds
 for $\rho R(k)\rho$, and
estimate B  holds by the Spectral Theorem.
The conclusion of Theorem 1 follows. The case of $\Gamma_-$ is
proven in the same way, using the obvious adaptation of
Lemma~\ref{8}. The theorem now follows from Lemma~\ref{8}

\end{section}
\begin{section}{Example of quasimode construction}
The following example is due to Buldyrev \cite{Bul}. Figure 2
shown below is the union of two circular arcs of radii $r_1$, $r_2$.
Under an assumption (see p.20 in \cite{Bul}) that will be satisfied for generic $r_1,r_2,d$,
Buldyrev then constructs a quasimode concentrated on the line segment
labelled $\ell$.
The associated frequencies are
$w_{p,q}^2$, with
$$w_{p,q}=\frac{1}{2d}\left ( \pi p+(q+\frac{1}{2})\arccos \sqrt
{(1-\frac{2d}{r_1})(1-\frac{2d}{r_2})}+O(\frac{1}{p}) \right ) ;$$
here $p,q$ are arbitrary positive integers. It is easy see that
for any $r_1,r_2,d$ and any fixed $q$, the sequence $w_{p,q}$ will
satisfy the asymptotics required in the hypothesis of Corollary 1.

It should be remarked that under weaker - and easier to verify - hypotheses
on $r_1,r_2,d$, Buldyrev's construction yields a sequence of
functions $u_j$ such that $\| (\Delta -\la_j)u_j\|_{L^2(\Omega )}
=O(\la_j^m)$ with $m<\infty$, and this would enable one to prove
a weaker version of Corollary 1.

Figure 4 shows one of the ways in which the circular arcs in figure
3 can be placed in a portion of a waveguide (actually only the portion
near the line segment $\ell$ is necessary for the quasimode construction).

\begin{figure}[ht]
\vspace{3.0in}
\noindent
\caption{
Circular arcs.
}

\vspace{-0.8in }

\vspace{0.6in }

\end{figure}

\begin{figure}[ht]
\vspace{3.0in}
\noindent
\caption{
Waveguide with resonances
}

\vspace{-0.8in }

\vspace{0.6in }

\end{figure}

\end{section}


\begin{thebibliography}{20}

\bibitem{apv} A. Aslanan, L. Parnovski, and D.G. Vassiliev,
``Resonances in acoustic waveguides'', Q. J. Mech. Appl. Math.
53 (3), p.429-447 (2000).

\bibitem{Bab} V.M. Babich and V.S. Buldyrev,
``Short-wavelength Diffraction Theory. Asymptotic Methods'',
Translated for the 1972 Russian original by E.F. Kuester.
Springer Series on Wave Phenomena, 4. Springer-Verlag,
Berlin, 1991.

\bibitem{Bul} B.S. Buldyrev, ``The asymptotic behaviour of the solutions
of the wave equation concentrated near the axis of  a two dimensional 
waveguide in an inhomogeneous medium'', in {\em Topics in Mathematical 
Physics, Volume 3: Spectral Theory}, ed. M. Sh. Birman, translated from
Russian, Consultants Bureau, New York 1969.

\bibitem{cart} M.L. Cartwright, {\em Integral Functions}, Cambridge
University Press, Cambridge UK (1962).

\bibitem{CZ} T. Christiansen and M. Zworski, ``Spectral asymptotics
for manifolds with cylindrical ends'', Ann. Inst. Fourier (1) 45 (1992)
251-267.

\bibitem{Do} H. Donnelly, ``Eigenvalue estimates for certain non-compact
manifolds'', Michigan Math. J 31 (1984), 349-357.


\bibitem{E} J. Edward, ``Eigenfunction decay and accumulation for the
Laplacian on asymptotically perturbed waveguides'', J. London Math. Soc. (2)
59 (1999) 620-636.

\bibitem{Va} D.V. EVANS, M.R. LEVITIN, and D.G. VASSILIEV,  ``
Existence theorems for trapped modes'', J. Fluid Mech. \underline{261}
(1994), 21-31.

\bibitem{GK} I. Gohberg and M. Krein, {\em Introduction to
the theory of linear nonselfadjoint operators}, Translations
of Mathematical Monographs,  (1969), AMS, Providence, RI.


\bibitem{Laz} V.F. Lazutkin, ``KAM theory and semiclassical
approximations to eigenfunctions''. With an addendum by
A.I. Schnirelman. Ergebnisse der Mathematik und ihrer
Grenzgebiete 24, Springer-Verlag, Berlin, 1993.

\bibitem{L} B.Ja. Levin, {\em Distribution of zeros of
entire functions}, AMS Translations of mathematical
monographs,  Vol. 5, 1980.

\bibitem{M} R.B. Melrose, ``Polynomial bounds on the distribution
of poles in scattering by an obstacle'', Journees ``Equations
aux derivees partielles'', Saint-Jean-de-Monts, 1984.


\bibitem{Mel} R.B. Melrose, {\em The Atiyah-Patodi-Singer index
theorem}, A.K. Peters, Wellesley, MA.. 1993.

\bibitem{P} L. Parnovski, ``Spectral asymptotics of the Laplace operator
on manifolds with cylindrical ends'', Internat. J. Math. 6 (1995) 911-920.

\bibitem{PZ} V. Petkov and M. Zworski, ``Breit-Wigner approximation and
the distribution of resonances'', Comm. Math.  Physics 204 (1999), no. 2,
329-351.

\bibitem{Ma} A.S. Markus, {\em Introduction to the spectral theory of 
polynomial operator pencils}, Translations of Mathematical Monographs, 
AMS Vol. 71, 1988, AMS, Providence, Rhode Island.

\bibitem{W2} K. MORGENROTHER and P. WERNER, ``On the principles
of
limiting absorption and limiting amplitude for a class of locally
perturbed waveguides. Part 1: time independant theory'', Math. Methods
Appl. Sci. \underline{10} (1988), 125-144.

\bibitem{W3} K. MORGENROTHER and P. WERNER, ``On the principles of
limiting absorption and limiting amplitude for a class of locally
perturbed waveguides. Part 2: time dependant theory'', Math. Methods
Appl. Sci. \underline{11} (1989), 1-25.


\bibitem{RS} M. REED and B. SIMON, {\em Methods of Modern Mathematical
Physics}, Vols. 1-4, New York, Academic Press, 1972.

\bibitem{shu} M. Shubin, {\em Pseudodifferential operators and
spectral theory}, Springer-Verlag 1987.

\bibitem{sj2} J. Sjostrand, ``Resonances for bottles and trace formulae'',
Math. Nachr. 211 (2001), 95-149.

\bibitem{Sj} J. Sjostrand, ``A trace formula and review of
some estimates for resonances'', in {\em Microlocal
analysis and spectral theory} (Lucca, 1996), 377-437,
NATO Adv. Sci. Inst. Ser. C Math. Phys. Sci., 490, Kluwer
Acad. Publ., Dordrecht, 1997.

\bibitem{SjZ} J. Sjostrand and M. Zworski, ``Complex scaling and
the distribution of scattering poles'', Journal of AMS 4,
(1991), 729-769.

\bibitem{SW} A. Soffer and M. Weinstein, ``Resonances, radiation
damping, and instability in Hamiltonian nonlinear wave equations'',
Invent. Math. 136 (1999), 9-74.

\bibitem{St} P. Stefanov, ``Quasimodes and resonances: sharp
 lower bounds'', Duke Math. J. 99 (1999), no. 1, 75-92.

\bibitem{st2} P. Stefanov, ``Resonances near the real axis
imply existence of quasimodes'', C.R. Acad. Sci. Paris Ser. 1 Math,
330 (2000), 105-108.

\bibitem{st3} P. Stefanov, ``Sharp upper bounds on the number of resonances
near the real axis for trapped systems'', preprint, (2001).

\bibitem{SV1} P. Stefanov and G. Vodev, ``
Distribution of resonances for Neumann problem in linear
elasticity outside a strictly convex body'', Duke Math. J. 78(3),
(1995), 677-714.

\bibitem{SV} P. Stefanov and G. Vodev, ``Neumann resonances
in linear elasticity for an arbitrary body'', Comm. Math. Phys. 176,
(1996), 645-659.

\bibitem{TZ} S. Tang and M. Zworski, ``From quasimodes to
resonances'', to be published in Math. Res. Letters.

\bibitem{Vain} B. Vainberg, {\em Asymptotic methods in equations of 
mathematical physics}, Gordon and Breach Science Publishers, New York, 1989.

\bibitem{V} G. Vodev, ``Sharp polynomial bounds on the number
of scattering poles for perturbations of the Laplacian'',
Comm. Math. Phys. 146 (1992), 205-216.

\bibitem{V2} G. Vodev, ``Sharp bounds on the number
of scattering poles in the two dimensional case'',
Math. Nachr. 170 (1994), 287-297.

\bibitem{SZ} G. Vodev, ``Sharp bounds on the number of
scattering poles in even-dimensional spaces'', Duke Math. J.
 74, (1994), 1-16.

\bibitem{V3} G.Vodev, ``Resonances in Euclidean scattering'',
preprint.

\bibitem{W} H. Weidenmuller, ``Studies of many-channel scattering'',
Ann. Physics 28 (1964), 60-115.

\bibitem{Z2} M. Zworski, ``Sharp polynomial bounds on the
number of scattering poles'', Duke. Math. J. 59, (1989), 311-323.

\bibitem{Z3} M. Zworski, unpublished, 1990.

\bibitem{Z} M. Zworski, ``Counting the scattering poles'',
Spectral and Scattering Theory, (M. Ikawa, ed.), Marcel Dekker,
1993.

\bibitem{Z9} M. Zworski, ``Poisson formula for resonances in even
dimensions'', Asian J. Math. 2, (1998), 609-617.



\bibitem{Z4} M. Zworski, ``Resonances in physics and geometry'',
Notices Amer. Math.  Soc. 46 (1999), 319-328.


\end{thebibliography}
\end{document}